\documentclass[%
reprint,
 amsmath,amssymb,
 aps,
]{revtex4-1}

\usepackage{graphicx}
\usepackage{dcolumn}
\usepackage{bm}
\usepackage[colorlinks,linkcolor=blue,anchorcolor=blue,citecolor=blue,urlcolor=black]%
{hyperref}
\begin{document}

\title{Quantum microwave-optical interface with nitrogen-vacancy centers in diamond}

\author{Bo Li}
\author{Peng-Bo Li}
\email{lipengbo@mail.xjtu.edu.cn}
\author{Yuan Zhou}
\author{Sheng-Li Ma}
\author{Fu-Li Li}
\affiliation{%
Shaanxi Province Key Laboratory of Quantum Information and Quantum
Optoelectronic Devices, Department of Applied Physics, Xi'an
Jiaotong University, Xi'an 710049, China
}%


\begin{abstract}
We propose an efficient scheme for a coherent quantum interface between microwave and optical
photons using nitrogen-vacancy (NV) centers in diamond. In this setup, an NV center ensemble
is simultaneously coupled to an optical and a microwave cavity. We show that, by using the
collective spin excitation modes as an intermediary, quantum
states can be transferred between the microwave cavity and the optical cavity  through either a
double-swap scheme or a dark-state protocol. This hybrid quantum interface may provide interesting
applications in single microwave photon detections or quantum information processing.

\end{abstract}

\maketitle
\section{introduction}

Microwave radiation is very commonly used both in our everyday life and in state-of-the-art
science and technology \cite{RMP-20-668}. As a practical technology, microwave radiation has
been widely applied to radar, communication, medical treatment and so on. In quantum science
and technology, microwave photons can be employed to couple solid state qubits such as
superconducting qubits \cite{RMP-85-623}. However, single-photon detection in the microwave
domain is extremely challenging, because microwave photon energies are in the milli-electron
volt range, 3 orders of magnitude smaller than in the visible or nearinfrared spectral
regions \cite{prl-114-113601}. On the other hand, many kinds of ultrasensitive detectors in
the optical frequency domain have been developed over the past decades. This implies that a
viable option for the detection of feeble microwave signals is via their conversion to the
optical frequency domain \cite{prl-114-113601,prl-116-023601,nc-3-1029}.

As for quantum information processing, different quantum systems may be combined in a hybrid device
for exploring new phenomenon or developing new quantum technology, which can make use of the best of
the components \cite{prl-97-033003,pra-87-144516,sr-6-33271,prl-117-015502,EPJ-2-18,npj-3-28,
oe-25-19226,pra-96-023827,LPL-14-025204,pra-88-042323,pra-85-022324,pra-94-043802}.
Generally speaking, microwave photons are perfect for manipulating
superconducting qubits or spins, while optical photons fit well for long distance transmissions.
Therefore, it is usually necessary to convert quantum states from microwave photons to optical photons
in the field of quantum information. As optical photons do not directly interact with microwave photons,
an intermediary is often required to exchange quantum information between them.

At present, there are several theoretical schemes or experimental works using different setups to realize
a quantum microwave-optical interface. For instance, proposals using one oscillator coupled to two cavities
with different wavelengths have been investigated theoretically \cite{pra-81-053806,prl-108-153603,prl-108-153604,
prl-109-130503,NSR-2-510} and experimentally \cite{natphys-9-712, natphys-10-321, natphys-507-81}, as nanomechanical
oscillators could couple to both microwave and optical photons via electro- and optomechanical forces respectively.
However, cooling the intermediate mechanical resonator to its ground state is still a great challenge currently.
Other schemes using conventional nonlinear crystals as a cavity electro-optic modulator have been proposed for
photonic conversion, but the quantum efficiency is still less than unity  \cite{pra-84-043845,prl-101-103601,
pra-92-043845}. Very recently, there are some proposals using NV centers in diamond or rare-earth-doped
crystals as an intermediary for a quantum photonic interface \cite{pra-91-033834, prl-113-063603,prl-113-203601,
pra-91-042307,pra-86-052325}. However, these schemes suffer from strong spin dissipations, in particular when a
spin ensemble is employed.

In this work, we consider a hybrid quantum interface using the collective spin excitations of an NV center
ensemble to transfer quantum states from microwave photons to optical photons. In particular,
the setup under consideration composes of an NV center ensemble, a microwave superconducting coplanar waveguide (CPW) cavity, and an optical cavity.
The NV spins are coupled to the microwave cavity via magnetic couplings, and simultaneously interact with the
optical cavity through an optical transition. We show that, in the low excitation limit, the collective spin excitations can be
mapped to a boson mode, and then could be used as a medium for the conversion. The system can be described by an
effective Hamiltonian composed of two Jaynes-Cummings (JC) interactions, one between the microwave cavity mode $\hat{a}_{1}$
and the collective spin mode $\hat{b}$, and the other between the optical cavity mode $\hat{a}_{2}$ and the
collective spin mode $\hat{b}$. Based on this effective interaction, we discuss two quantum state conversion protocols,
i.e., a double-swap protocol and a dark-state scheme. Different from the previous works \cite{pra-91-033834, prl-113-063603,
prl-113-203601,pra-91-042307,pra-86-052325}, here we introduce a dark mode of the collective
spin excitations, and propose an adiabatic conversion approach with this dark mode. We show that  the conversion
process is extremely robust against spin dissipations, as the dark mode is decoupled from the collective spin excitations.

\begin{figure*}
\includegraphics[width=12.0cm]{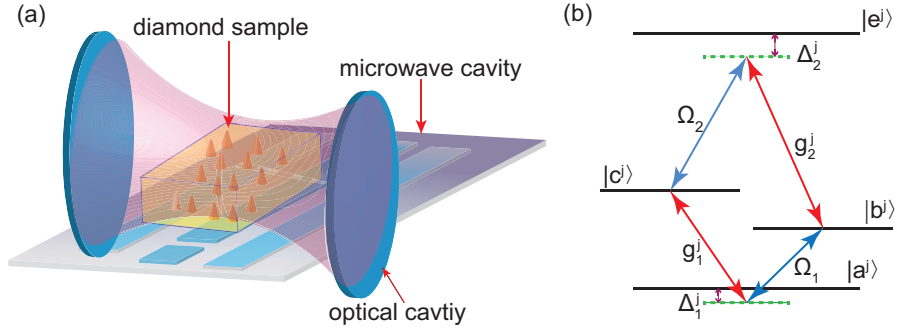}
\caption{\label{fig_setup}(Color online) Schematic design and operation for the micro-optical
interface. (a) Schematic of the device. An optical cavity with embedded NV centers is placed above a
CPW cavity. (b) Level diagram describing the interactions between the $j$th spin, the
CPW resonator and the optical cavity. Each spin is modeled as a four-level system, with two classical fields
$\Omega _{1}$ and $\Omega _{2}$ driving dispersively the transitions $|b^{j}\rangle \leftrightarrow |a^{j}\rangle $
and $|e^{j}\rangle \leftrightarrow |c^{j}\rangle $. The CPW resonator and optical cavity modes couple the transitions
$|c^{j}\rangle \leftrightarrow |a^{j}\rangle $ and $|e^{j}\rangle \leftrightarrow |b^{j}\rangle $.}
\end{figure*}

\section{The setup}

As shown in Fig.~1(a), the hybrid quantum device under consideration consists
of an ensemble of NV centers, a CPW resonator, and an optical cavity. Moreover, two
coherent driving fields and an extra static magnetic field are applied. The NV centers are
coupled to the CPW resonator, optical cavity and two external driving fields
simultaneously,  which forms a four-level system as illustrated in Fig.~1(b).

An NV center in diamond consists of a substitutional nitrogen atom replacing a
carbon atom and an adjacent vacancy, having trapped an additional electron.
The electronic ground state of the NV center is a spin-1 triplet, denoted as
$|^{3}A_{2}\rangle =|E_{0}\rangle \otimes |m_{s}=0,\pm 1\rangle $, with
$|E_{0}\rangle $ labels the orbital state with zero angular momentum
projection along the N-V axis. The resonance transition frequency between the degenerate sublevels
$|m_{s}=\pm 1\rangle $  and $|m_{s}=0\rangle $  is $2\pi \times 2.87$ GHz.
In our scheme, an extra static magnetic field $B_{0}$ is
applied to remove the degeneracy of the states $%
|m_{s}=1\rangle $ and $|m_{s}=-1\rangle $, with a Zeeman splitting $g_{e}\mu _{B}B_{0}$,
where $g_{e}=2$ is the NV land\'{e} factor, and $\mu _{B}=14\text{MHz}$ $\text{mT}^{-1}$ is
the Bohr magneton. We label the ground states as
$|a^{j}\rangle =|E_{0}\rangle \otimes |m_{s}=0\rangle $,
$|b^{j}\rangle=|E_{0}\rangle \otimes |m_{s}=-1\rangle $, and
$|c^{j}\rangle =|E_{0}\rangle \otimes|m_{s}=1\rangle $, while
the excited optical state $|e^{j}\rangle $ is chosen as
$|A_{2}\rangle =\frac{1}{\sqrt{2}}(|E_{-}\rangle \otimes |m_{s}=+1\rangle
+|E_{+}\rangle \otimes |m_{s}=-1\rangle )$
\cite{natrue-466-730,pra-83-054306,prl-97-247401,prl-101-117601},
where $|E_{\pm }\rangle $ denote the orbital states with
angular momentum projection $\pm 1$ along the N-V axis. The non-uniform strain may
affect the orbital spin level composition, and the optical transition selection rules and polarization properties. However, under the low strain condition, i.e., the non-axial crystal strain is much smaller than the spin-orbit splitting, the optical excited state $|A_{2}\rangle$ can still be available.

The interaction of the system can be described by two Raman transitions.
The frequencies for the CPW resonator, optical cavity and two classical fields
are $\nu _{1}$, $\nu _{2}$, $\omega _{1}$, and $\omega _{2}$ respectively. As displayed in Fig.~1(b),
the CPW resonator and optical cavity couple the transitions $|c^{j}\rangle \leftrightarrow |a^{j}\rangle $
and $|e^{j}\rangle \leftrightarrow |b^{j}\rangle $, while  two classical fields drive dispersively the transitions
$|b^{j}\rangle \leftrightarrow |a^{j}\rangle $ and $|e^{j}\rangle \leftrightarrow |c^{j}\rangle $, with Rabi
frequencies $\Omega _{1}$ and $\Omega _{2}$. The detunings for these transitions are
$-\Delta _{1}^{j}=\omega
_{ba}^{j}-\omega _{1}=\omega _{ca}^{j}-\nu _{1}$, $\Delta _{2}^{j}=\omega
_{ec}^{j}-\omega _{2}=\omega _{eb}^{j}-\nu _{2}$.
Due to the fluctuating magnetic environment, there might be fluctuations
in the detuning of levels $|b^j\rangle$ and $|c^j\rangle$. However, these environmental induced
fluctuations are much smaller than the frequency detunings, and can be safely ignored.
The $j$th spin located at $r_{j}$ is coupled to the
two cavities with the coupling strengths $g_{1}^{j}\propto B_{1}(r_{j})$, and
$g_{2}^{j}\propto E_{2}(r_{j})$, where $B_{1}(r_{j})$, and $E_{2}(r_{j})$ are the zero-point
magnetic and electric fields of the cavity modes $1$ and $2$, respectively.
As the collective enhanced couplings are employed, here we introduce $g_{1}=\sqrt{\frac{1}{N}\sum_{j=1}^{N}|g_{1}^{j}(r_{j})|^{2}}$, and $g_{2}=\sqrt{\frac{1}{N}\sum_{j=1}^{N}|g_{2}^{j}(r_{j})|^{2}}$ to denote the average coupling strengths
for each spin \cite{prl-107-060502,prl-107-220501,prl-103-070502}.
Then in the interaction picture, the Hamiltonian
of the system under the dipole and rotating wave approximation reads (let $\hslash=1$)
\begin{eqnarray}
H_{I} &=&\hat{a}_{1}\sum_{j=1}^{N}g_{1}|c^{j}\rangle \langle
a^{j}|e^{i\Delta _{1}^{j}t}+\Omega _{1}\sum_{j=1}^{N}|b^{j}\rangle \langle
a^{j}|e^{i\Delta _{1}^{j}t}+  \nonumber \\
&&\hat{a}_{2}\sum_{j=1}^{N}g_{2}|e^{j}\rangle \langle b^{j}|e^{i\Delta
_{2}^{j}t}+\Omega _{2}\sum_{j=1}^{N}|e^{j}\rangle \langle c^{j}|e^{i\Delta
_{2}^{j}t}  \nonumber \\
&&+H.c.,
\end{eqnarray}
where $\hat{a}_{i}$ is the
annihilation operator for the cavity $i(i=1,2)$.

The presence of random local strain may lead to
inhomogeneous broadening in the transition frequencies, and then results in random shifts
$\delta _{1}^{j}=\Delta _{1}^{j}-\Delta _{1}$,and $\delta _{2}^{j}=\Delta _{2}^{j}-\Delta
_{2}$  for the $j$th spin, where $\Delta_{1}$, and $\Delta_{2}$ are the average detunings.
Here we consider the system under the large detuning condition, i.e.,
$|\Delta _{1}|\gg |\Omega _{1}|,|g_{1}|,|\delta _{1}^{j}|$, and $|\Delta _{2}|\gg|\Omega_{2}|,|g_{2}|,|\delta _{2}^{j}|$,
and ignore the inhomogeneous broadening of the transition frequencies in the following.
In this case, the states $|e^{j}\rangle $ and $|a^{j}\rangle$ could be adiabatically eliminated, as they dispersively couple
to the states $|b^{j}\rangle $ and $|c^{j}\rangle $. Then we obtain the effective Hamiltonian
\cite{fp-48-823,cjp-85-625,prl-118-083604,prl-113-023603, pra-77-015809} describing the hybrid system
\begin{eqnarray}
H_{eff} &=&(\frac{|\Omega _{1}|^{2}}{\Delta _{1}}+\frac{|g_{2}|^{2}}{\Delta
_{2}}\hat{a}_{2}^{\dagger }\hat{a}_{2})\hat{J}_{bb}+(\frac{|\Omega _{2}|^{2}%
}{\Delta _{2}}+\frac{|g_{1}|^{2}}{\Delta _{1}}\hat{a}%
_{1}\hat{a}_{1}^{\dagger })\hat{J}_{cc}  \nonumber \\
&&+(\frac{\Omega _{1}g_{1}^{\ast }}{\Delta _{1}}\hat{a}_{1}^{\dagger }+\frac{%
\Omega _{2}g_{2}^{\ast }}{\Delta _{2}}\hat{a}_{2}^{\dagger })\hat{J}_{bc}+(%
\frac{\Omega _{1}^{\ast }g_{1}}{\Delta _{1}}\hat{a}_{1}+\frac{\Omega
_{2}^{\ast }g_{2}}{\Delta _{2}}\hat{a}_{2})\hat{J}_{cb},
\end{eqnarray}
with $\hat{J}_{mn}=\sum_{j=1}^{N}|m^{j}\rangle \langle n^{j}|$.
We will ignore the first two terms corresponding to nearly homogeneous energy shifts for each spin,
as they could be compensated by tuning the frequencies of the cavities and the classical fields.
The last two terms describe the two cavities couple to the collective electron spin wave excitations of NV centers.

In the low excitation limit, we could map the collective spin operators $\hat{J}_{cb}(\hat{J}_{bc})$
into boson operators $\hat{b}^{\dagger }(\hat{b})$ by introducing the Holstein--Primakoff representation
\begin{eqnarray}
\nonumber
\hat{J}_{cb}=\hat{b}^{\dagger }\sqrt{N-\hat{b}^{\dagger
}\hat{b}}\simeq \sqrt{N}\hat{b}^{\dagger }
\\
\hat{J}_{bc}=\hat{b}\sqrt{%
N-\hat{b}^{\dagger }\hat{b}}\simeq \sqrt{N}\hat{b}
\nonumber\\
\hat{J}_{z}=(\hat{%
b}^{\dagger }\hat{b}-\frac{N}{2}),
\end{eqnarray}
where the operators $\hat{b}$ and $%
\hat{b}^{\dagger }$ approximately obey the standard boson commutator $[\hat{b}$, $\hat{b}^{\dagger
}]=1$ \cite{prl-103-070502,prl-105-210501,pra-88-013837}.
Then, the effective Hamiltonian is given by
\begin{equation}
H_{eff}=G_{1}(t)\hat{a}_{1}\hat{b}^{\dagger }+G_{2}(t)\hat{a}_{2}\hat{b}^{\dagger }+H.c.,
\end{equation}%
with $G _{1}=\frac{\Omega _{1}^{\ast }\sqrt{N}g_{1}}{\Delta _{1}}$, and $G _{2}=\frac{%
\Omega _{2}^{\ast }\sqrt{N}g_{2}}{\Delta _{2}}$, corresponding to  the effective collective
coupling strengths. These effective couplings can be dynamically controlled by the Rabi frequencies $\Omega _{i}$
and detunings $\Delta _{i}$. In the following, we assume that $\Omega _{1}$, $\Omega _{2}$,
$g_{1}$ and $g_{2}$ are real for simplicity.

Since any quantum system would suffer from decoherence, here we consider the
decay rates $\kappa _{1}$, $\gamma _{s}$, and $\kappa _{2}$ for the CPW resonator,
collective spin mode, and optical cavity respectively. We assume the setup works in the low-temperature environment,
the thermal photon occupation numbers is nearly zero, i.e., $n_{1,2}=(e^{\hslash \nu _{1,2}/k_{B}T}-1)^{-1}\simeq 0$.
Then the dynamics of the
system can be described by the following master equation
\begin{equation}
\frac{d\hat{\rho}}{dt}=-i[H_{eff},\hat{\rho}]+\kappa _{1}\mathcal{D}[\hat{a}%
_{1}]\hat{\rho}+\kappa _{2}\mathcal{D}[\hat{a}_{2}]\hat{\rho}+\gamma _{s}%
\mathcal{D}[\hat{b}]\hat{\rho},
\end{equation}
where $\mathcal{D}[\hat{o}]\hat{\rho}=\hat{o}\hat{\rho}\hat{o}^{\dag }-\frac{1}{2}%
\hat{o}^{\dag }\hat{o}\hat{\rho}-\frac{1}{2}\hat{\rho}\hat{o}^{\dag }\hat{o}$
for a given operator $\hat{o}$.

The effective Hamiltonian of the system in Eq. (4) describes two JC interactions, one between
the microwave cavity $\hat{a}_{1}$ and the collective spin excitation mode $\hat{b}$,
the other between $\hat{b}$ and the optical cavity $\hat{a}_{2}$. This beam-splitter Hamiltonian is analogous to a mechanical
resonator coupled to two electromagnetic cavities \cite{pra-81-053806,prl-108-153603}. We would take the collective
spin mode as a medium to exchange quantum states between the two cavities. In what follows, we will consider
two different protocols, i.e., the double-swap conversion and dark-state conversion schemes.

\section{Double-swap conversion}

The JC model is initially developed to describe the interaction between a
two-level atom and a single-mode field \cite{QO-1997, PIEEE-51-89}. In that case, the photons keep
oscillating between the two states. Utilizing the dynamics of the system,
population transfer between the two levels can be realized. Similarly, JC
interactions between two boson modes can be used to transfer quantum states.
In our scheme, the collective spin mode $\hat{b}$ interacts with two cavity modes, respectively.
Therefore, it is not difficult to use the spin mode to swap the quantum
states between the microwave mode and the optical mode.

Generally, the double-swap protocol includes three steps: step 1, prepare the spins to their
ground state in time $0<t<T_{0}$, which could be realized by the optical pumping method;
step 2, turn on the coupling $G _{1}$ (while $G _{2}=0$) in time $T_{0}<t<T_{1}$,
to transfer quantum states from the microwave mode $\hat{a}_{1}$ to the collective spin mode $\hat{b}$;
step 3, turn off $G _{1}$ and turn on the coupling  $G _{2}$  in time $T_{1}<t<T_{2}$, to
map quantum states from the spin mode $\hat{b}$ to the optical mode $\hat{a}_{2}$.

\begin{figure}
\includegraphics[width=9.4cm]{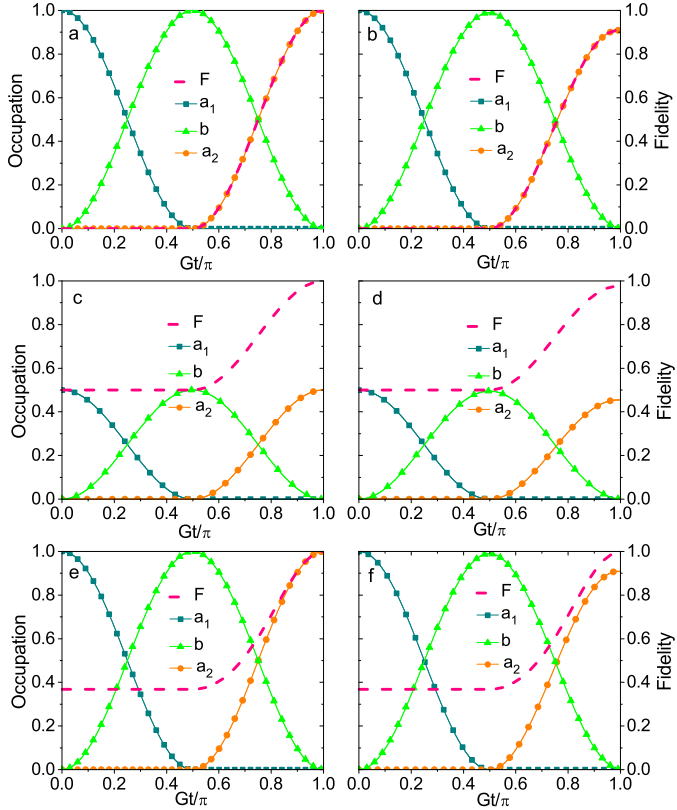}
\caption{\label{fig_double}(Color online) Fidelity ($F$) and occupations ($\hat{a}_{1},\hat{b},\hat{a}_{2}$) as
a function of time in the double-swap conversion process,
with the coupling parameters $G_{1}=G_{2}=G$.
Three kinds of initial states are under consideration:
(\romannumeral1). a Fock state $|1\rangle$ in (a) and (b); (\romannumeral2). a superposition state
$\frac{1}{\sqrt{2}}|0\rangle +\frac{1}{\sqrt{2}}|1\rangle $ in (c) and (d);
(\romannumeral3). a coherent state $|\alpha\rangle $, $\alpha=1$  in (e) and (f).
The decay parameters for (a), (c), and (e) are chosen as
$\protect\kappa _{1}=\protect\gamma _{s}=\protect\kappa _{2}=0$, and
for (b), (d), and (f) $\protect\kappa _{1}=0.003G,\protect\gamma _{s}=0.01G,\protect\kappa _{2}=0.1G$.}
\end{figure}

In this protocol, the effective Hamiltonian of the system reads
\begin{equation}
H_{eff}=
\left\{
\begin{array}{ll}
G _{1}(\hat{a}_{1}\hat{b}^{\dagger}+\hat{a}_{1}^{\dagger}\hat{b}) \ \ \   T_{0}<t<T_{1}, \\
G _{2}(\hat{a}_{2}\hat{b}^{\dagger}+\hat{a}_{2}^{\dagger}\hat{b}) \ \ \   T_{1}<t<T_{2}.
\end{array}
\right.
\end{equation}
By solving the Heisenberg equations in the period from $T_{0}$ to $T_{1}$, the dynamics
of the operators can be derived as
\begin{eqnarray}
\nonumber
\hat{a}_{1}(t)=\cos (G _{1}t)\hat{a}_{1}-i\sin (G _{1}t)\hat{b},\nonumber\\
\hat{b}(t)=\cos (G _{1}t)\hat{b}-i\sin (G _{1}t)\hat{a}_{1}.
\end{eqnarray}
When $t=\frac{\pi }{2G _{1}}$, we have $%
a_{1}(t)=-ib$ and $b(t)=-ia_{1},$ corresponding to a complete exchange of quantum states
between the microwave mode $\hat{a}_{1}$ and the collective spin mode $\hat{b}$ except a phase factor $e^{i\frac{3}{2}\pi }$.
In the period from $T_{1}$ to $T_{2}$, the quantum states could be exchanged in the same way between the spin
mode $\hat{b}$ and the optical mode $\hat{a}_{2}$.

The conversion efficiency can be evaluated by Uhlmann fidelity \cite{RMP-9-273}, which is
defined as $F=(Tr[(\sqrt{\rho _{1}}\rho _{2}\sqrt{\rho _{1}})^{1/2}])^{2}$,
with $\rho _{1}(\rho _{2})$ the density matrix of cavity 1(2) at the beginning(end) of the transfer process.
Then, we perform numerical simulations for the double-swap protocol by solving the master equation (5) with the effective Hamiltonian (4), where the coupling parameters are $G_{1}=G_{2}=G$,
and $G \sim 2\pi \times 1$ MHz as discussed in Sec. \uppercase\expandafter{\romannumeral5}.
We prepare the optical cavity and collective spin mode to their ground states,
and study the effects of the decay parameters on the fidelity of the protocol with different initial states.
Three kinds of initial states for the microwave cavity are under consideration: a Fock state $|1\rangle$ as displayed in Fig.~2(a) and (b),
a superposition state $\frac{1}{\sqrt{2}}|0\rangle +\frac{1}{\sqrt{2}}|1\rangle$ as displayed in Fig.~2(c) and (d),
and a coherent state $|\alpha\rangle,\alpha=1$ as displayed in Fig.~2(e) and (f).

We first consider the conversion in the ideal
case in which the decoherence processes are neglected. The results show that a fidelity as high as $1.0$ can be reached
for all initial states, as displayed in  Fig.~2(a),(c),(e). We find that, as time evolves, the quantum
state of the microwave cavity is transferred to the collective spin mode completely when $Gt=\frac{\pi }{2}$,
and finally to the optical cavity when $Gt=\pi$. Then, when it comes to the realistic case, we consider the
dissipation processes with the decay rates $\kappa _{1}=0.003G$ for the microwave cavity, $\gamma _{s}=0.01G$
for the collective spin mode, and $\kappa _{2}=0.1G$ for the optical cavity.
Although the decoherence processes have a harmful effect on the conversion process,
high fidelities 0.90,0.97,0.99 can still be reached  for the initial quantum states $|1\rangle$,
$\frac{1}{\sqrt{2}}|0\rangle +\frac{1}{\sqrt{2}}|1\rangle$ and $|\alpha\rangle$ respectively, as shown in Fig.~2(b),(d),(f).
Therefore, the double-swap protocol works very well under the realistic conditions.

\section{dark-state conversion}

Using the probe and pump pulses in a counterintuitive sequence, the well-known STIRAP technique
has become an established procedure for coherent population transfer in a three-level system \cite{pra-52-566}.
In this adiabatic process, as the conversion is preserved in a ``dark" dressed state---one
eigenstate of the system with zero eigenvalue in the interaction picture---the atomic
spontaneous decay could be effectively suppressed \cite{pra-52-566,pra-52-583,pra-71-3095}.
Further study shows similar adiabatic protocols are applicable to other physical systems as well,
such as an optomechanics system \cite{prl-108-153604,prl-108-153603}, or a hybrid quantum device
\cite{prapl-04-044003}. Here, we consider an adiabatic dark-state transfer protocol based upon this model.

\begin{figure}
\includegraphics[width=9.4cm]{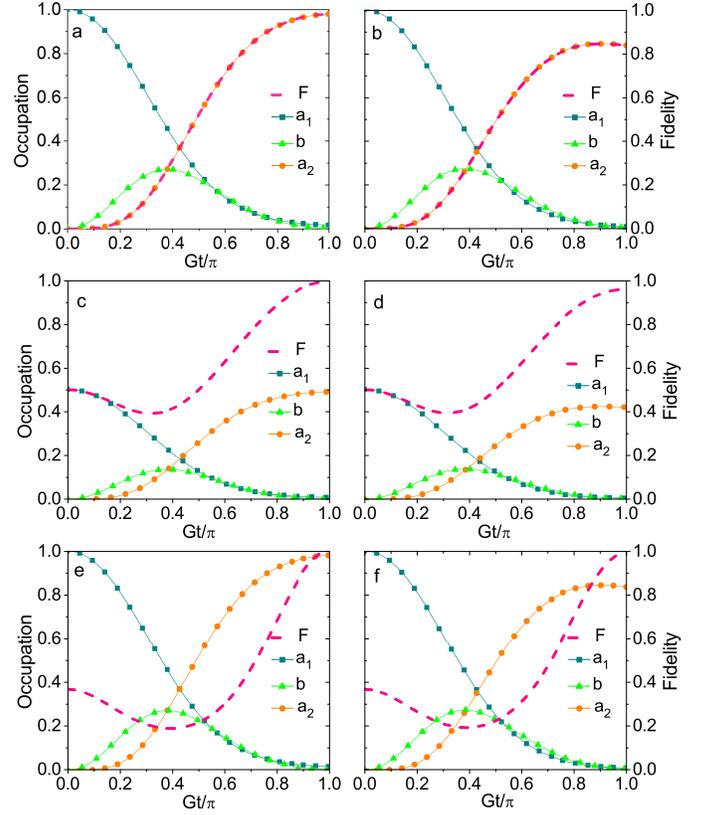}
\caption{\label{fig_dark_1}(Color online) Fidelity ($F$) and occupations ($\hat{a}_{1},\hat{b},\hat{a}_{2}$) as
a function of time in the dark-state scheme,
with the coupling parameters $G_{1}(t)=Ge^{\frac{(t-2.8)^{2}}{20}}$ and $G_{2}(t)=1.45Ge^{\frac{-t^{2}}{6}}$.
Three kinds of initial states are under consideration:
(\romannumeral1). a Fock state $|1\rangle$ in (a) and (b); (\romannumeral2). a superposition state
$\frac{1}{\sqrt{2}}|0\rangle +\frac{1}{\sqrt{2}}|1\rangle $ in (c) and (d);
(\romannumeral3). a coherent state $|\alpha\rangle $, $\alpha=1$  in (e) and (f).
The decay parameters for (a), (c), and (e) are chosen as
$\protect\kappa _{1}=\protect\gamma _{s}=\protect\kappa _{2}=0$, and
for (b), (d), and (f) $\protect\kappa _{1}=0.003G,\protect\gamma _{s}=0.01G,\protect\kappa _{2}=0.1G$.}
\end{figure}

As to our system, mapping the spin excitations to a
boson mode, the free Hamiltonian of the system reads $H_{0}=\nu _{1}\hat{a}_{1}^{\dagger }%
\hat{a}_{1}+\nu _{s}\hat{b}^{\dagger }\hat{b}+\nu _{2}\hat{a}_{2}^{\dagger }\hat{a}_{2}$, with
$\nu _{s}$ the frequency of the collective spin mode. When the
interaction Hamiltonian given by Eq. (4) is taken into consideration, the eigenmodes of the
system will change into hybridized forms, as discussed below in details.

We introduce two hybridized boson modes describing
quasiparticles formed by combinations of microwave and optical photons as $%
\hat{c}_{d}=-\cos \theta \hat{a}_{1}+\sin \theta \hat{a}_{2}$, $\hat{c}%
_{b}=\sin \theta \hat{a}_{1}+\cos \theta \hat{a}_{2}$, with $\tan \theta =G_{1}/G_{2}$.
Then, to describe quasiparticles hybridized with microwave, optical photons and spin
excitations, two other boson modes are introduced $\hat{c}_{\pm }=(1/\sqrt{2%
})(\hat{c}_{b}\pm \hat{b})$. It can be readily verified that the Hamiltonian
of the system can take the form as $H_{0}+H_{eff}\approx \omega _{d}\hat{c}%
_{d}^{\dagger }\hat{c}_{d}+\omega _{+}\hat{c}_{+}^{\dagger }\hat{c}%
_{+}+\omega _{-}\hat{c}_{-}^{\dagger }\hat{c}_{-}$, with $\omega
_{d}=\nu _{s}$, $\omega _{\pm}=\nu _{s}\pm \sqrt{\nu _{1}^{2}+\nu _{2}^{2}}$. The hybridized
modes $\hat{c}_{d}$, $\hat{c}_{+}$, $\hat{c}_{-}$ are the eigenmodes of the
system, distinguished by different eigen energies.

We refer to $\hat{c}_{d}$ as a spin dark mode, as it only involves the cavity
modes, decoupled from the collective spin excitations. Similar to the unpopulated intermediate
level in the STIRAP process, the collective spin excitation mode remains unaffected when
the hybridized eigenmode $\hat{c}_{d}$ is excited. Especially, in the limit $%
\theta =0$, we get $\hat{c}_{d}=-\hat{a}_{1}$, while in the limit $\theta =\pi /2$, we get $\hat{c}_{d}=%
\hat{a}_{2}$. This implies, if we adiabatically rotate the mixing angle $%
\theta $ from $0$ to $\pi /2$, the spin dark  mode $\hat{c}_{d}$
would evolve from $-\hat{a}_{1}$ to $\hat{a}_{2}$. Utilizing this feature,
the quantum states of one cavity could be converted to the other via the collective
spin excitation mode but without actually populating it.

The adiabatic dark-state protocol is similar to the well-known STIRAP scheme. We modulate
the coupling strengths $G_{1}(t)$ and $G_{2}(t)$ so that the spin dark mode $\hat{c}_{d}$  adiabatically
evolves from being $-\hat{a}_{1}$ at the beginning to $\hat{a}_{2}$ at the end. As a result, the quantum state
of the microwave cavity $\hat{a}_{1}$ would be transferred to the optical cavity $\hat{a}_{2}$ finally. To keep the conversion process in the spin dark-state, the coupling strengths should be varied slowly in order to maintain the adiabatic conditions. At the same time, in consideration of the decoherence of the system, the transfer process should be finished before the dissipations seriously affect the process.

The numerical results for the dark-state protocol can be obtained by solving the master equation  (5) with the effective Hamiltonian (4),  where the coupling parameters $G_{1}(t)=Ge^{\frac{(t-2.8)^{2}}{20}}$, $G_{2}(t)=1.45Ge^{\frac{-t^{2}}{6}}$, and $G \sim 2\pi\times 1$ MHz as discussed in
Sec. \uppercase\expandafter{\romannumeral5}. It should be noted that these coupling parameters could further be optimized .
The optical cavity and collective spin mode are prepared to their ground states.
To test the robustness of the scheme, we consider three kinds of initial states for the microwave cavity:
a Fock state $|1\rangle$ as displayed in Fig.~3(a) and (b), a superposition state $\frac{1}{\sqrt{%
2}}|0\rangle +\frac{1}{\sqrt{2}}|1\rangle$ as displayed in Fig.~3(c) and (d), and a coherent state
$|\alpha\rangle,\alpha=1$ as displayed in Fig.~3(e) and (f).

We now discuss the dark-state protocol in the ideal and practical cases.
In the ideal case without decoherence,  a fidelity as high as $0.99$ can be reached for all the initial states, as displayed in  Fig.~3(a),(c), and (e).
We find that, as the system evolves, the quantum state of the microwave
cavity is slowly transferred to the optical cavity in the
conversion process. Different from the double-swap scheme, the collective spin mode only has few excitations in the conversion process.
When it turns to the practical case, we take the decay parameters as $\kappa _{1}=0.003G,\gamma_{s}=0.01G,\kappa _{2}=0.1G$, and show that
maximum fidelities $0.84, 0.95$, and $ 0.99$ can be reached for the initial quantum states $|n\rangle = 1$,
$\frac{1}{\sqrt{2}}|0\rangle +\frac{1}{\sqrt{2}}|1\rangle$ and $|\alpha\rangle$ respectively, as shown in  Fig.~3(b), (d), and (f).

To study the dependence of the fidelity on the collective spin decay rates,
we simulate the dark-state protocol with different damping parameters, as displayed in Fig.~4.
We choose a superposition state $\frac{1}{\sqrt{2}}|0\rangle +\frac{1}{\sqrt{2}}|1\rangle$ as the initial state, and employ
the same coupling parameters as those in Fig.~3. The decay rates of the cavities are $\kappa _{1}=0.003G$ and $\kappa _{2}=0.1G$.
We show that, when increasing the collective spin decay rate $\gamma _{s}$
from 0.01G to 0.1G, the fidelity decreases in a very limited range. This verifies that the dark-state protocol
is extremely robust against the spin dissipations, as the spin dark mode is decoupled from the spin excitation modes.

\begin{figure}
\includegraphics[width=9.0cm]{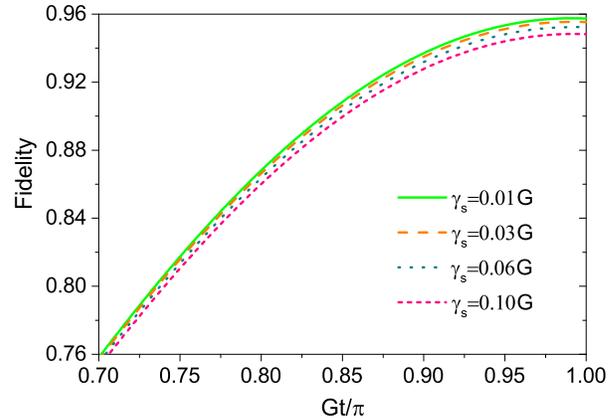}
\caption{\label{fig_dark_2}(Color online) Fidelity as a function of time in the dark-state
conversion scheme with different  spin decay rates: $\gamma _{s}=0.01G$ (green solid curve),
$\gamma _{s}=0.03G$ (orange dashed curve), $\gamma _{s}=0.06G$ (dark cyan dotted curve), and $\gamma _{s}=0.1G$ (pink short-dashed curve).
The initial state is chosen as a superposition state $\frac{1}{\sqrt{2}}|0\rangle +\frac{1}{\sqrt{2}}|1\rangle$,
with the same coupling parameters as those in Fig. 3. The decay parameters for both cavities are
$\protect\kappa _{1}=0.003G$, and $ \protect\kappa _{2}=0.1G$.}
\end{figure}

\section{EXPERIMENTAL FEASIBILITY}

We now discuss the experimental implementation of our scheme. The present-day achievements in the
experiment with NV centers coupled to microwave (optical) cavities could be utilized.
The strong interaction between NV centers and a CPW cavity has been experimentally demonstrated \cite{
prl-105-140502,prl-107-220501,prl-107-060502,prl-105-140501}. In our case, we
consider an ensemble with $1\times 10^{12}$ NV centers, which would generate
a collective coupling constant $\sqrt{N}g_{1} \sim 2\pi \times 10$ MHz. At the temperature of
$T \sim 20$ mK, the equilibrium thermal photon occupation numbers are less than 0.01, then could be neglected.
As for the optical coupling,
our scheme can be realized with several kinds of optical cavities, such as whispering-gallery-modes (WGM)
in microsphere (microdisk) resonators, or Fabry-P\'{e}rot cavities with high
quality factors. Strong  interactions between individual NV centers in diamond and WGM in
a microsphere (microdisk) resonator have been reached, with coupling strengths
$2\pi \times (0.3-1)$ GHz \cite{NL-6-2075,NL-9-1447,NL-8-3911,JPB-42-114001,JPB-95-191115,
NT-21-274008,OE-17-8081,pra-85-042306,OE-23-13734}. The coupling strength between
NV centers and an optical cavity could be modulated by the position of the NV ensemble near the cavity.
Here we take $\sqrt{N}g_{2} \sim 2\pi \times 500$ MHz. Considering the inhomogeneous broadening in the transition frequencies are about
$\delta _{1}^{j} \sim 2\pi \times 10$ MHz \cite{prl-113-023603}, $\delta _{2}^{j} \sim 2\pi \times 10$ GHz  \cite{NT-21-274008}, we choose the
microwave detuning $\Delta _{1} \sim 2\pi \times 200$ MHz, and optical detuning $\Delta _{2} \sim 2\pi \times 100$ GHz.
We further assume the laser Rabi frequencies $\Omega _{1} \sim 2\pi \times 20$ MHz, and $\Omega_{2} \sim 2\pi \times 200$ MHz.
Then we obtain the effective Raman transition rates $G_{1} = G_{2}= G \sim 2\pi \times 1$ MHz.

In the practical situations, a quality factor  $Q\sim 10^{6}$ for the CPW cavity is realistic \cite{RMP-85-623},
which would lead to a decay rate $\kappa _{1} \sim 2\pi \times 3$ kHz. Besides, a coherence time
longer than 100 $\mu s$ for an NV center ensemble has been demonstrated \cite{prb-82-201201},
corresponding to $\gamma _{s} \sim 2\pi \times 10$ kHz.
The inhomogeneous broadening caused by nitrogen electronic spins or $^{13}C$ nuclear
spins may limit the long spin-coherence time of NV spins. However, it can
be compensated by narrowing of the nuclear field distribution or the spin-echo techniques \cite{prl-110-250503,prx-5-031031}, which will prolong the desphasing time from $T_2^*$ to $T_2$ \cite{prl-110-250503,prx-5-031031}.
As for the optical cavity, even challenging, a high quality
factor $Q\sim 10^{9}$ can still be reached \cite{RMP-85-623,natrue-443-671}, then the photon decay rate
could be estimated as $\kappa _{2} \sim 2\pi \times 100$ kHz. As discussed in the above sections,
the numerical simulations displayed in Fig.~2, Fig.~3 and Fig.~4 are performed with these coupling (decay) parameters.

\section{Conclusion and discussion}

In summary, we have presented an efficient scheme for a micro-optical
interface, which would be potentially used in microwave photon detections or quantum
information conversions. Our device composes of an NV center ensemble,
coupled to a CPW cavity and an optical cavity, respectively. In the low excitation limit,
the collective excitations of the spin ensemble could be mapped to a boson
mode, and mediates the quantum state mapping between the two cavities.
Then the effective Hamiltonian of the system can be performed as a beam-splitter form.
Based upon this model, we discuss a double-swap protocol and a dark-state scheme for quantum state conversions.
For the dark-state transfer protocol, we show that one eigenmode of the system is the spin dark
mode, which is decoupled from the collective spin excitations. Modulating the coupling parameters properly under the
adiabatic conditions, the quantum state of one cavity could be transferred to
the other with very high fidelities. As the conversion process is kept evolving in the spin dark-state,
the decay of the NV spins could be effectively suppressed.

Quantum states conversion at the subphoton level is an attractive subject to explore.
In quantum technology, it means switching the low energy signal or quantum information
from one degree to another. With this hybrid quantum device, both the double-swap protocol
and the dark-state scheme are available. Here the dark-state scheme
is particularly robust against spin dissipations.
This hybrid quantum device may offer a realistic quantum micro-optical interface.

Since cold atoms have the advantage of negligible inhomogeneous broadening in the transition frequencies,
this scheme can be applied to cold atomic ensembles as well. Given that the energy level structure
of cold atoms may be different, this scheme can also be implemented in a three-level
system (see the Appendix for more details). For both the three- or four-level models, the collective
coupling  between the spin (atomic) ensemble and the microwave cavity  plays a central role
for the photonic conversion. As for the cold neutral atoms, the strong coupling of an ultracold gas to
a CPW resonator has been demonstrated \cite{prl-103-043603,RMP-85-623,pra-79-040304}.
Compared to solid-state systems like NV centers, although the
technology of trapping a cold atomic gas in an optical cavity has been achieved \cite{RMP-85-553}, it
may still complicate the experimental realization.

\section*{Acknowledgments}

This work is supported by the NSFC under
Grant Nos. 11774285 and 11474227, and the Fundamental
Research Funds for the Central Universities. Part of the
simulations are coded in PYTHON using the QUTIP library \cite{CPC}.

\section*{APPENDIX}

The core idea of this manuscript employs the collective excitation modes of the spin ensemble as an intermediary for
microwave-optical interface. We give the analysis based upon a four-level system in the above main text.
In what follows, we show that this scheme can be implemented in a three-level system as well. Further more, taking the cold $^{87}Rb$ atoms
as an example, we will make a detail discussion.

Here we consider a hybrid quantum device utilizing an ensemble of cold $^{87}Rb$ atoms
coupled to the two cavity modes simultaneously. In this case, utilizing the two cavity modes and a classical
field coupled to the cold atomic ensemble, we could establish a three-level system and implement both the double-swap
protocol and dark-state scheme. Specifically, we choose the ground states as $|b\rangle =|5^{2}S_{1/2},F=1\rangle $,
$|c\rangle =|5^{2}S_{1/2},F=2\rangle $, and choose the excited optical state as $|e\rangle
=|5^{2}P_{3/2},F=2\rangle $; then a three-level system is established as displayed in Fig.~5.

\begin{figure}[tbp]
\includegraphics[width=5.50cm]{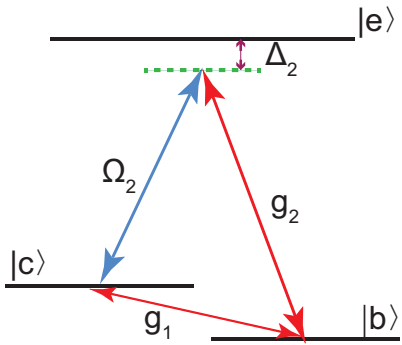}
\caption{(Color online) Level diagram describing the
interactions between the cold atomic ensemble and the CPW resonator as well as the optical cavity.
Each atom is modeled as a three-level system, with
the classical fields $\Omega _{2}$ driving  the transitions $|e\rangle \leftrightarrow
|c\rangle $. The two cavity modes couple the
transitions $|c\rangle \leftrightarrow |b\rangle $ and $|e\rangle
\leftrightarrow |b\rangle $.}
\label{fig_setup}
\end{figure}

In the interaction picture, the Hamiltonian of the system
under the  rotating wave approximation reads
\begin{eqnarray}
H_{I} &=&\hat{a}_{2}\sum_{j=1}^{N}g_{2}|e^{j}\rangle \langle
b^{j}|e^{i\Delta _{2}^{j}t}+\Omega _{2}\sum_{j=1}^{N}|e^{j}\rangle \langle
c^{j}|e^{i\Delta _{2}^{j}t}  \nonumber \\
&&+\hat{a}_{1}\sum_{j=1}^{N}g_{1}|c^{j}\rangle \langle b^{j}|+H.c..
\end{eqnarray}
Similar to the previous analysis in Sec. \uppercase\expandafter{\romannumeral2},
we adiabatically eliminate the level $|e\rangle $ in the large detuning conditions,
and map the collective spin operators into boson mode. Then we can obtain the effective
Hamiltonian describing the system
\begin{equation}
H_{eff}=G_{1}^{^{\prime }}(t)\hat{a}_{1}\hat{b}^{\dagger }+G_{2}(t)\hat{a}_{2}\hat{b}^{\dagger }+H.c.,
\end{equation}%
where $G_{1}^{^{\prime }}=\sqrt{N}g_{1}$, and $G_{2}=\frac{\Omega _{2}^{\ast }%
\sqrt{N}g_{2}}{\Delta _{2}}$. This beam-splitter Hamiltonian is similar to the result
of four-level system in Eq. (4), while the only difference is between $G_{1}^{^{\prime }}$ and $G_{1}$.

Note that this model can apply to other three-level system as well, such as an ensemble of NV centers or an erbium-doped crystal.
The main difference between the three- and four-level systems is the effective collective
coupling strength between the ensemble and the microwave cavity. One may choose either of them for the implementation of this scheme.

\end{document}